# Improved visualisation of brain arteriovenous malformations using color intensity projections with hue cycling

Keith S Cover
VU University Medical Center
Amsterdam, The Netherlands

## Abstract

Color intensity projections (CIP) have been shown to improve the visualisation of greyscale angiography images by combining greyscale images into a single color image. A key property of the combined CIP image is the encoding of the arrival time information from greyscale images into the hue of the color in the CIP image. A few minor improvements to the calculation of the CIP image are introduced that substantially improve the quality of the visualisation. One improvement is interpolating of the greyscale image s in time before calculation of the CIP image. A second is the use of hue cycling – where the hue of the color is cycled through more than once in an image.  The hue cycling allows the variation of the hue to be concentrated in structures of interest. If there is a zero time point hue cycling can be applied after zero time and before zero time can be indicated by greyscale. If there is an end time point hue cycling can be applied before the end time and pixels can be set to black after the end time. An angiogram of a brain is used to demonstrate the substantial improvements hue cycling brings to CIP images. A third improvement is the use of maximum intensity projection for 2D rendering of a 3D CIP image volume. A fourth improvement allowing interpreters to interactively adjust the phase of the hue via standard contrast- brightness controls using lookup tables. Other potential applications of CIP are also mentioned.

**Introduction**

Cover et al. (2006; 2007A) proposed a method for combining any number of grey scale images into a single color image. In color intensity projections (CIP), when a pixel has the same intensity in all the component images it has the same grey scale value in the combined image. However, if the intensity of the pixel changes over the images the combined pixel has a hue that encodes the time of the maximum pixel intensity. The algorithm presented herein includes minor improvements on Cover et al. (2006; 2007B) that lead to substantial improvements in the CIP image. These minor improvements include hue cycling.

In hue, one complete cycle is red-yellow-green-light blue-blue-purple-red. Cover et al. (2006, 2007A, 2007B) only used red through blue so that each hue would map to a unique time. However, there are situations, such as arrival time CIP in astronomy and angiography, were rapidly varying hue may be more useful than being able to map the hue to a unique time (Cover 2012). To test this improvement, the current paper compares cycling hue several times over an image in angiogram to no hue cycling.

While other orders of the hue - other than red-yellow-green-light blue-purple-red - are of course possible, the ordering used in the current paper is the one most widely used.

CIP was originally developed for applications in medical imaging (Cover et al.  2006, 2007A), such as angiography and MRI including MRI perfusion, but the generality of the technique has made it applicable to many other applications.  Other applications include astronomy (Cover et al. 2007; Cover 2012) and visualisation flow including air, other gases, liquids and solids (Merzkirch, 1987; Settles, 2001; Smits & Lim, 2000).

**Method**

For comparison purposes, the angiogram used to demonstrate the improvements to CIP is the same as the one used in Cover et al. (2006). The selection of the greyscale images for the angiogram can also be found in the same publication.  The patient was a 46-year-old man who experienced epilepsy as a result of a Spetzler-Martin Grade II arteriovenous malformation (AVM) feeding from the middle cerebral artery in the left temporal lobe of the brain.

Before the interpolation the brightness of the grey scale images was reversed.  The reversal resulted in the background color changing from white to black and the presence of contrast to cause pixels to become brighter rather than darker.  The reversal was implemented by first finding the maximum of all the pixels in the 29 greyscale images. Each of the pixel values in the 29 greyscale images was then subtracted from the maximum to get the new value to the reversed value of the pixel.

Before calculation of the CIP, the 29 greyscale images were interpolated in time to 113 images. The interpolation was only in time, no spatial interpolation was employed.

Two different CIP images were calculated from the 113 greyscale images one without hue cycling and one with. The CIP without hue cycling cycled through the full range of hues over 99 interpolated greyscale images. A value less than 113 was used because several of the early and late greyscale images had little information.

For hue cycling the hue was cycled over 22 of the interpolated greyscale images.  The number 22 was arrived at by trying a range of values and choosing the one that gave the most informative image. A smaller number than 22 yielded a large number of artefacts. A larger number did not present all the information available with hue cycling.

The brightness, hue and saturation of the color of each pixel of the CIP image was calculated individually from the greyscale images on a pixel by pixel basis.  The range of hue and saturation is typically zero to one. The CIP image was calculated by the following simple equations

Brightness = MaxIP ,
Saturation = ( MaxIP - MinIP ) /MaxIP                    (1)
Hue =  (index of the image of the maximum value for the pixel)/(images per hue cycle)

Where MaxIP is the maximum value of the pixel over the greyscale images and MinIP is the minimum value of the pixel over the greyscale images. The image index is 0 for the first image, 1 for the second image, and 113 for the last image. For example, if a pixel had the maximum value in the first image, with no hue cycling its hue would be red. For the last image the hue would be purple.

To display the CIP hue-saturation-brightness image it is usually necessary to convert it to an RGB image. While there are a variety of different ways to accomplish this conversion the algorithm used for this paper was the HSB to RGB conversion routine in the Java Programming Language.

The saturation can also be amplified by multiplying the saturation by a constant after it has been calculated with the above equation. Scaling the saturation can give a better indication of how much the intensity of a pixel has varied over the component greyscale images. For a moving object on a black background the change in intensity is always 100% so the saturation is also 100%. Also, in digital subtraction angiography the saturation is always 100% as the first image is before the contrast arrives and is black. However, for some images the saturation may be much smaller and there may be a desire to amplify it.

**Results and Discussion**

Figure 1 show the CIP for the angiogram of the example AVM with interpolation but without hue cycling. The order of the hue with increasing time is red-yellow-green-light blue-blue. The contrast entering the image is shown in red while the contrast leaving the image is shown in blue.

The black background, due to the inversion of the brightness in the greyscale image before calculation of the CIP, improves the contrast of the image. The interpolation of the greyscale images before also improves the contrast.

The AVM is prominent in the centre of the image with most of its vessels shown in yellow. The artery feeding the AVM are orange/yellow while the ones leaving it are green. Close examination of Figure 1 shows that each hue in the image only corresponds to a single arrival time as expected for no hue cycling.

Figure 2 is calculated in the same way as Figure 1 with the exception that the hue is cycled in the image. For example, the purple hue at the bottom of the image indicates both early and late arrival times.

Examination of the hues in the AVM show red vessels feeding it with the body of the AVM made of orange and yellow vessels. As with Figure 1 the draining vessels are green. The increased range of hues as compared to Figure 1, due to hue cycling, makes for better differentiation of the vessels in the AVM.

Another way to deal with a large number of times encoded as hues in an image, other than hue cycling, is to limit the images to just a few consecutive hues – such as red-yellow-green-light blue-blue – but allow the user to move the hues back and forward in time interactively. If there are more than 5 time points in an image, 5 consecutive time points can be assigned red-yellow-green-light blue-blue. Additionally, any time points before red can be assigned red and any time points after blue can be assigned blue.

However, the use of a limited number of hues requires the user to be able to interactively move the hues forward and back in time. Most image display applications allow the adjustment of the brightness/contrast of the image. Many also allow the loading of lookup tables (LUTs) that map greyscale values to RGB colours. By combining the two it is possible for a user to control the phase of the hue and thus effectively move back and forth in time.

The first stage is to encode both the brightness and time of each pixel in a greyscale value. For example, the lower 6 bits can be used represent time points from 0 to 63 in monotonically increasing order. The next higher 8 bits encode a brightness between 0 and 255.

The next stage is to set up a greyscale to RGB LUT that works in steps of 64 levels and with 256 steps for a total of 16384 levels. Each of the 256 steps maps to a unique brightness between 0 and 255 and are ordered by increasing brightness.

The 64 levels in each step map to 64 hues. The first 30 levels map to a hue of red. The 31, 32, 33, and 34 levels map to hues of yellow, green, light blue and blue. The 35 to 63 levels map to a hue of blue.

Loading an image display application with both the images and LUTs just described will perform as desired provided the brightness and contrast are adjusted accordingly. The contrast, which is a scale factor, should be held constant at a value of unity ensuring each increase of one level increase of the "brightness" slider controlled by the user, which is an offset, yields a one level increase in the LUT.

As an example, consider a pixel that has a brightness of 218 and a time point of 32. Initially – with the "brightness" slider at 0 and the contrast at unity – the LUT will map the pixel to a fairly bright yellow. However, when the user increases the "brightness" slider by exactly 1 level while keeping the contrast unchanged at unity, the pixel will have the same brightness, but the hue will be moved up from yellow to green.

Therefore, as long as the "brightness" slider is kept within a limited range, say –10 to 10 levels, and the contrast fixed at unity, the "brightness" slider will allow the user to move the hue back and forward in time.

Consequently, in some cases, no special software is required to enable users to move the hue of CIP back and forward in time – just carefully designed LUTs.

In some cases there may be a clear zero time point where only the timing of the pixels after the zero time is of interest. In this case pixels before the zero time can have their saturation set to 0% rather than the 100% for hue cycling. Thus the pixels before the zero time point will have the brightness of hue cycling pixels but not the timing and will be greyscale. Thus a radiologist reviewing the image will see greyscale pixels before the zero time and hue cycled pixels after the zero time.

In some cases there may be a clear end time beyond which the value of the pixels is uninformative or unhelpful. In these cases the color of the pixels after the end time can be set to black. Black pixels do not visually interfere with the color pixels before the end time.

In cases were the angiography is acquired in 3D it is straightforward to calculate a corresponding 3D CIP. A common practice for rending 3D greyscale image volumes in 2D is the maximum intensity projection (MIP). Wikipedia defines MIP as a volume rendering method for 3D data that projects in the visualisation plane the voxels with maximum intensity that fall in the way of parallel rays traced from the viewpoint to the plane of projection. To generalise MIP from a 3D greyscale image volume to a 3D CIP, instead of projecting the pixel with the maximum intensity in each ray, the pixel with the maximum brightness is projected in each ray, along with its hue and saturation. The resulting 2D CIP rendering will behave similarly to MIPs in other ways including the rotation of the MIPs.

**Conclusions**

The introduction of interpolation, hue cycling and maximum intensity projections have improved the contrast of the vessels in the AVM. The option of allow control of the phase of the hue via the standard brightness/contrast adjustment can be useful. These improvements may be useful in other applications of CIP.

**Acknowledgements** This work was funded by the VU University Medical C enter in Amsterdam.  Disclosure: The authors employer, the VU University Medical Center in Amsterdam, is pursuing patents on color intensity projections covering applications in astronomy, medical imaging, and other fields. The author has a financial interest in the patent.

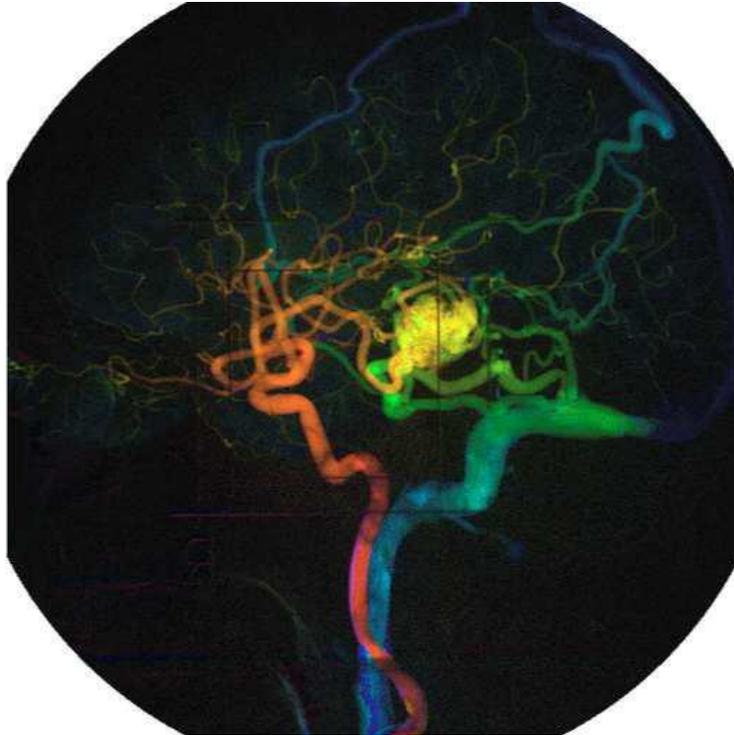

*Figure 1. CIP image with interpolation but without hue cycling*

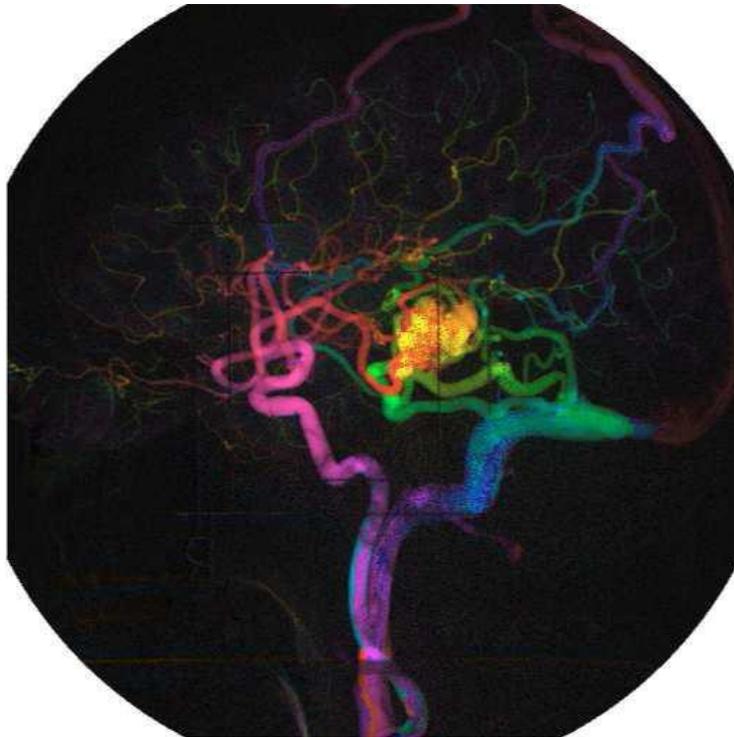

**F**igure 2. CIP image with interpolation but with hue cycling